\documentclass[aps,pre,twocolumn,letterpaper,floatfix,showpacs]{revtex4-1}
\usepackage{graphicx}
\usepackage{amsmath,amssymb,amsfonts}
\usepackage{mathtools}
\usepackage[hidelinks]{hyperref}
\usepackage{epstopdf}

\begin{document}

\title{A minimal model for the onset of slip pulses in frictional rupture}

\author{Kjetil Th\o gersen$^\text{a}$}
\email{kjetil.thogersen@fys.uio.no}
\author{Einat Aharonov$^\text{b}$}
\author{Fabian Barras$^{a}$}
\author{Fran\c{c}ois Renard$^\text{a,c}$}
\affiliation{
The Njord Centre, Departments of Physics and Geosciences, University of Oslo, 0316 Oslo, Norway$^\text{a}$ \\
Institute of Earth Sciences, The Hebrew University, Jerusalem, 91904, Israel$^\text{b}$\\
Universit{\'e} Grenoble Alpes, Universit{\'e} Savoie Mont Blanc, CNRS, IRD, IFSTTAR, ISTerre, 38000 Grenoble, France$^\text{c}$}
\date{\today}

\begin{abstract}
We present a minimal one-dimensional continuum model for the transition from crack-like to pulse-like propagation of frictional rupture. In its non-dimensional form, the model depends on only two free parameters: the non-dimensional pre-stress and an elasticity ratio that accounts for the finite height of the system. The model predicts stable slip pulse solutions for slip boundary conditions, and unstable slip pulse solutions for stress boundary conditions. Results demonstrate that the existence of pulse-like ruptures requires only elastic relaxation and redistribution of initial pre-stress. The novelty of our findings is that pulse-like propagation along frictional interfaces is a generic elastic feature, whose existence does not require a particular rate- or slip-dependencies of dynamic friction.
\end{abstract}

\maketitle

\section{Introduction}
Frictional rupture, the process by which relative sliding motion starts along the interface between two contacting solid surfaces, is of prime importance in the description of various systems in physical sciences and engineering. Examples range from the squealing of brake pads and bush bearings \cite{kinkaid2003automotive,chen2019effect}, corrugation and wear of mechanical compounds \cite{kato2000wear}, to the earthquake cycle along crustal faults \cite{scholz1998earthquakes}, and the surge of glaciers \cite{thogersen2019rate}.
A frictional rupture is characterized by its rupture speed and its rupture mode. The rupture speed can vary from slow, through sub-Rayleigh up to super-shear \cite{thogersen2019minimal,ben2010dynamics,tromborg2014slow}. The rupture mode can be described by analogy to the dynamics of shear cracks, a behavior referred to as \textit{crack-like} dynamics \cite{svetlizky2014classical}. Frictional rupture can also be \textit{pulse-like} rupture, where the slipping portion of the interface is spatially and temporally constrained to the vicinity of the propagating rupture tip.

Since the seminal work of \citet{heaton1990evidence} in 1990, the pulse-like rupture mode has been successfully used to rationalize the short slip duration observed in seismic inversions of earthquakes and describe the propagation of both fast \cite{chen2020cascading,galetzka2015slip} and slow \cite{michel2019similar} earthquakes. Pulse-like ruptures have also been reproduced in laboratory experiments \cite{baumberger2002self,shlomai2016structure,shlomai2020onset,Lu2010pulse,lykotrafitis2006self}. The ubiquity of pulse-like rupture observed in frictional systems has motivated the development of theoretical and numerical models to investigate the conditions supporting the emergence of slip pulses. 
Slip-pulses have been found in systems with a large variety of boundary conditions and domain approximations \cite{carlson1989mechanical,carlson1994dynamics,elbanna2012new, brener2018unstable,zheng1998conditions,brener2018unstable,brantut2019stability,nielsen2003self,gabriel2012transition,putelat2017phase,erickson2011periodicity,gerde2001friction} and friction laws including rate-and-state friction \cite{zheng1998conditions,gabriel2012transition,brener2018unstable,putelat2017phase,erickson2011periodicity}, velocity-weakening friction \cite{carlson1989mechanical,carlson1994dynamics,perrin1995self}, slip-dependent friction \cite{wu1998solitary} and Coulomb friction \cite{nielsen2003self}.

Several studies have demonstrated the important control of the initial stress distribution along the interface, also called pre-stress, on slip pulses. Pulses widen with increasing pre-stress, approaching crack-like rupture \cite{Lu2010pulse,melgar2017systematic}. Heterogeneous stress distributions can cause slip pulses \cite{day1998dynamic}. Stress barriers can cause cracks to arrest and create a backward propagating arresting front that causes a pulse in the opposite direction \cite{johnson1990initiation}. Also, fault geometry may play a role. Large earthquakes have been shown to favour a transition from crack-like to pulse-like propagation \cite{melgar2017systematic}. Large aspect ratios (i.e. large earthquakes where rupture length is larger than other length scales such as seismogenic depth) may favour the formation of slip pulses \cite{nielsen2000influence}, and the seismogenic depth can limit the width of earthquake slip pulses \cite{ampuero2017upper,bai2017effect}. Studies have also demonstrated the inherent instability of slip-pulses for different friction constitutive laws \cite{gabriel2012transition,brener2018unstable,noda2009earthquake,brantut2019stability}.

Despite these extensive studies, the fundamental set of conditions required to develop slip pulse in frictional systems remains debated. Several studies have highlighted the need of velocity-weakening friction to obtain slip-pulses, and there is a growing consensus that geometrical effects and initial pre-stress may play a role, but that the existence of slip-pulses requires either strong velocity-weakening friction \cite{shlomai2020onset,chen2020cascading,Lu2010pulse,beeler1996self} or a dependency of normal stress on local slip due to, for example, bi-material effects \cite{shlomai2020onset} or thermal pressurization of pore fluids along the frictional interface \cite{garagash2012seismic}.

In light of the many different models and interpretations of slip pulses, our goal is to identify the minimum ingredients necessary for their appearance. Here, we introduce a one-dimensional continuum model which, in non-dimensional form, contains only two free parameters; the non-dimensional initial shear pre-stress, $\bar \tau$, and a ratio of elastic moduli, $\bar \gamma$, that governs the elastic relaxation of shear stress due to a finite system size. We demonstrate that the onset of slip pulses can be explained in full by the elastic relaxation or redistribution of initial pre-stress. In our simplified model, the existence of slip pulses does not rely on a dependency of normal stress on slip, nor velocity weakening friction. The stability of the pulse-like propagation is set by the boundaries. For stress boundary conditions, pulses are sensitive to small perturbations in the pre-stress along the interface, which control their shape. For displacement boundary conditions, the model predicts a transition from crack-like propagation to steady-state pulses for constant pre-stress.

The manuscript is structured as follows. We derive the non-dimensional equation of motion of a minimal friction model in section \ref{sec:simplified_model}, with supporting equations in appendix \ref{app:eom} and \ref{app:eom_nondim}. We then demonstrate how this minimal model contains a transition from crack-like to pulse like propagation for two different choices of boundary conditions in section \ref{sec:crack-pulse}, and derive scaling relations for this this transition. We then discuss the results in context of existing literature on slip pulses in section \ref{sec:discussion}, before we sum up and conclude in section \ref{sec:concusion}. Additional information and equations can be found in the appendices, which contain more detailed derivations of the equations of motion (appendix \ref{app:eom} and \ref{app:eom_nondim}), analytical solution of slip pulses (appendix \ref{app:analytical}), as well as additional figures investigating the role of viscous damping on the numerical results (\ref{app:scaling_damping}).

\section{Simplified model of pulse-like rupture \label{sec:simplified_model}}
\begin{figure}
    \centering
    \includegraphics[width = .49\textwidth]{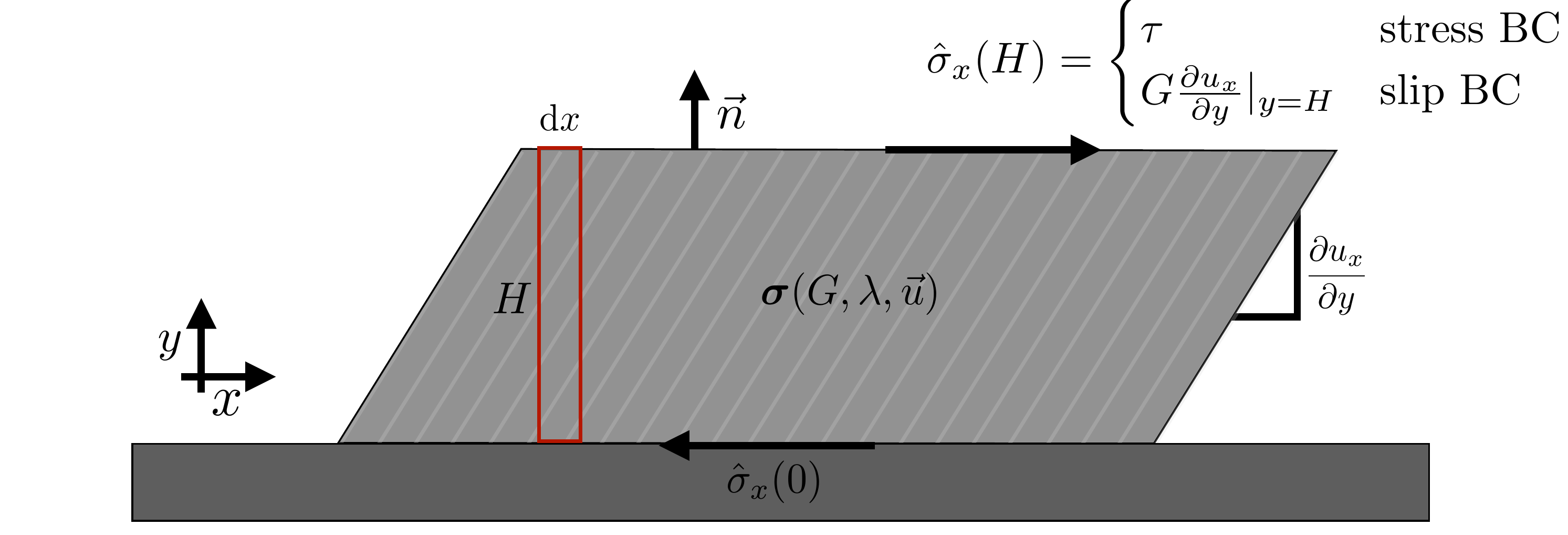}
    \caption{Sketch of the two-dimensional system that is integrated to obtain the one-dimensional equation of motion used in the manuscript. We model a thin elastic layer of thickness $H$ with shear modulus $G$ and the second Lamé coefficient $\lambda$. Two boundary conditions (BC) are considered on the top surface. At $y=H$ we apply either a constant stress $\tau$ or a fixed slip $\partial u_x/ \partial t = 0$ which can be translated to a stress boundary condition through linearization of $u_x(y)$. At $y=0$ we apply an Amontons-Coulomb friction law. The system is integrated across the y-coordinate (red rectangle) to obtain a one-dimensional approximation.}
    \label{fig:BC_sketch}
\end{figure}
We solve one-dimensional elasticity by integrating over a thin layer with thickness $H$, lying above an infinitely rigid substrate. Here, $H$ is the only length scale of the model and this length scale needs to be much smaller than any other length scales involved in frictional rupture, such as rupture length or slip pulse width. This approximation is useful in the sense that the system is primed for frictional ruptures that propagate over distances large compared to the system thickness $H$, conditions that observations have shown to favour slip pulses. Note that a similar approach has been used previously by e.g. \citet{bouchbinder2011slow}. Here, we use two types of boundary conditions at the top surface, i) displacement boundary condition and ii) stress boundary condition. 

We start with the two-dimensional problem presented in the FIG. \ref{fig:BC_sketch}, following the approach of \citet{bouchbinder2011slow}. We consider a long linear elastic layer with height $H$ in contact with an infinitely rigid substrate. $H$ is assumed to be much smaller than any length scale over which all other quantities vary in the x-direction. The equation of motion is
\begin{equation}
    \rho \frac{\partial^2 \vec u}{\partial t^2} = \nabla \cdot \boldsymbol{\sigma},
\end{equation}
where the stress tensor $\boldsymbol{\sigma}$ is given by
\begin{equation}
    \boldsymbol{\sigma} = \lambda \text{tr}(\nabla \vec u) \boldsymbol{I} + G [\nabla \vec u + (\nabla \vec u)^T].
\end{equation}
$G$ is the shear modulus, $\lambda$ is the second Lam{\'e} coefficient, and $\vec u$ is the displacement. In this 2D system, the boundary conditions are described in \figurename~\ref{fig:BC_sketch}. At the bottom interface, the shear stress is set by the frictional force $f_f$. On the top surface, two kinds of boundary conditions are considered, either imposed shear stress or imposed displacement. To obtain a one-dimensional approximation, we integrate the equation of motion over a finite thickness $H$, corresponding to the red rectangle in \figurename~\ref{fig:BC_sketch}, and solve the momentum equation for the average horizontal displacement $\langle u_x \rangle_y$. Next, we assume the normal stress $\sigma_n$ to be independent of $x$ and $\dot{u}_y =0$, which prevents a dependency of normal stress on slip. For constant $G$ and $\lambda$, we obtain (detailed derivation can be found in appendix \ref{app:eom}):
\begin{equation}
H \rho \frac{\partial^2 \langle u_x \rangle_y}{\partial t^2} = H (\lambda + 2G)\frac{\partial^2 \langle u_x \rangle_y}{\partial x^2} +  \hat \sigma_{x}(x,H,t) - \hat \sigma_x(x,0,t).
\end{equation}
The boundary conditions are set through the surface tractions $\hat \sigma_{x}(x,H,t)$ and $\hat \sigma_x(x,0,t)$. At the $y=0$, we assume Amontons-Coulomb friction:
\begin{align}
    \hat \sigma_x(x,0,t) = f_f 
    \begin{cases}
        \leq \mu_s \sigma_n & \text{if } \dot u = 0 \\
        = \mu_k \sigma_n & \text{if } |\dot u| > 0
    \end{cases}
    \label{eq:ff}
\end{align}
where $\mu_s$ and $\mu_k$ are the static and dynamic friction coefficients. Next, we use two variations of boundary condition at $y=H$, constant displacement or constant stress.

\subsection{Non-dimensional formulation}
In non-dimensional form, the two variations of boundary condition at $y=H$ can be written as a single equation of motion (appendix \ref{app:eom_nondim}). Dropping the $\langle \rangle$ and the subscript $_x$, the momentum equation in the slipping portion of the interface for Amontons-Coulomb friction reduces to
\begin{equation}
\ddot{\bar u} = \frac{\partial^2 \bar u}{\partial \bar x^2} - \Gamma \bar \gamma {\bar u} + \bar \tau^\pm - \bar \beta \frac{\partial ^2 \dot {\bar u}}{\partial \bar x^2}.
\label{eq:eom}
\end{equation}
where $\Gamma$ is a binary variable that selects the boundary condition at the top interface
\begin{align}
\Gamma = 
\begin{cases}
0, & \text{if stress boundary condition}\\
1, & \text{if displacement boundary condition}.\\
\end{cases}
\end{align}
The non-dimensional pre-stress is 
\begin{equation}
    \bar \tau^\pm = \frac{ \tau/\sigma_n \mp \mu_\text{k}}{\mu_\text{s} - \mu_\text{k}}
\end{equation}
where $\tau$ is the initial shear stress along the interface. $\pm$ corresponds to the sign of the velocity. In this paper only positive velocities appear, which means we can simplify to $\bar \tau \equiv \bar \tau^+$. $\bar \tau$ is a function which sets the non-dimensional initial shear stress along the frictional interface. The static friction threshold in non-dimensional form can be written as $|\frac{\partial^2 \bar u}{\partial \bar x^2} - \Gamma \bar \gamma \bar u + \bar \tau| \geq 1$.
\begin{equation}
\bar \gamma = \frac{2G}{\lambda + 2G}
\end{equation}
is a dimensionless ratio of elastic parameters. $\bar x = x/X$, $\bar u = (u-u_0)/U$, $\bar t = t/T$ are the dimensionless position, slip, and time respectively. The characteristic length and time scales are chosen so that a non-dimensional front speed $\bar v_c = 1$ corresponds to the wave speed of the material; $X = H$, $U = 2 H \frac{\mu_\text{s} p - \mu_\text{k} p}{\lambda + 2G}$ (or $U = H \frac{\mu_\text{s} p - \mu_\text{k} p}{\lambda + 2G}$), $T = H \sqrt{\frac{\rho}{\lambda + 2G}}$. To limit oscillations on the scale of the discretization, we also introduce a small damping term $-\bar \beta \frac{\partial ^2 \dot {\bar u}}{\partial \bar x^2}$ where we use $\bar \beta = 10^{-3}$. The transition from dynamic to static friction occurs when the local slip velocity reaches zero. We solve the system for varying $\bar \tau (\bar x)$ and $\bar \gamma$. It is worth noting that the finite difference discretization of equation \ref{eq:eom} is identical to the dimensionless form of the classical Burridge-Knopoff spring block model under the additional constraint that the spatial discretization $\text{d}x$ equals the thickness of the system $\text{d}x = H$ ($\text{d} \bar x = 1$).

\begin{figure}
\includegraphics[width = .49\textwidth]{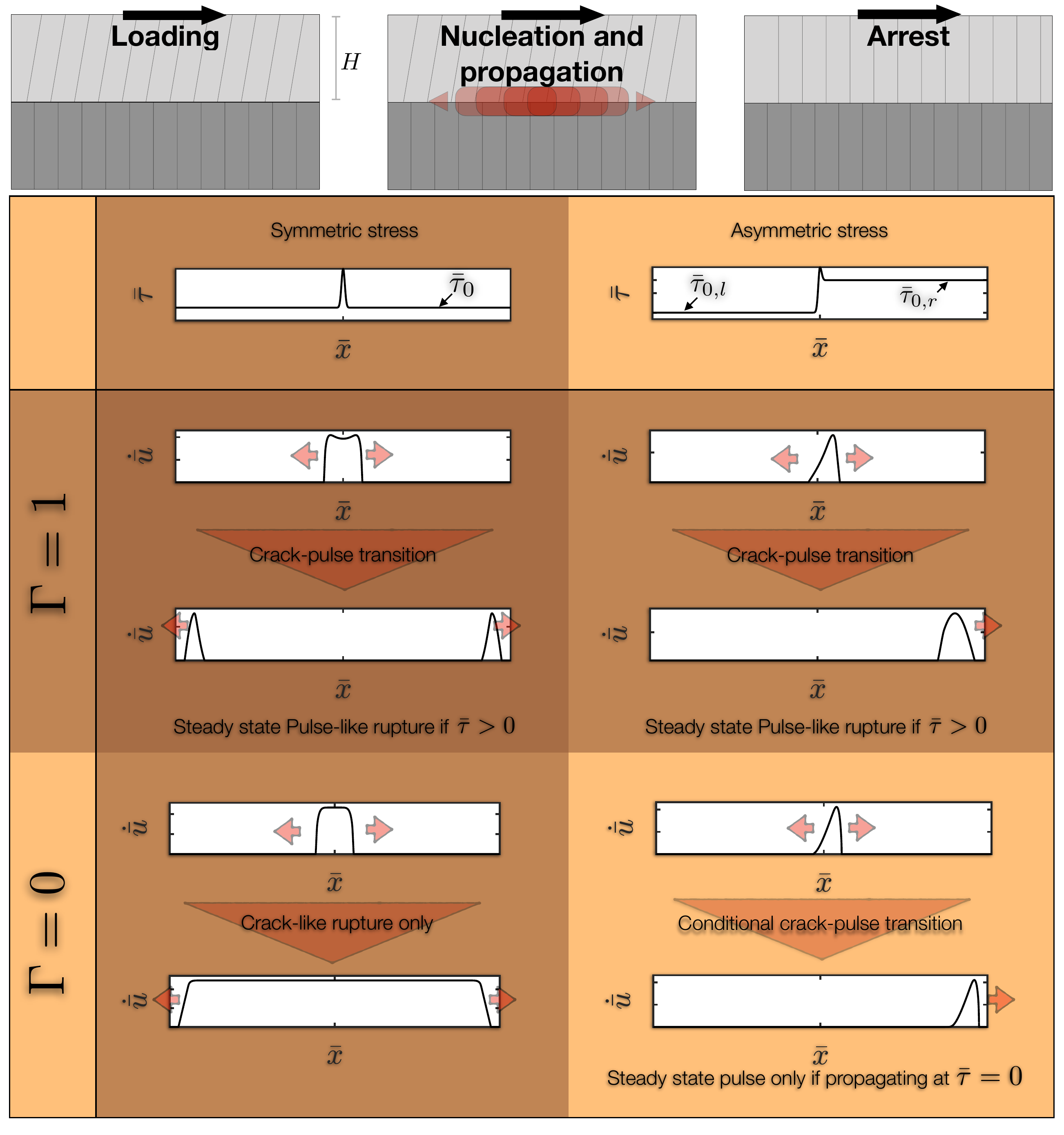}
\caption{Phase diagram of the behavior of crack-like versus pulse-like rupture in the parameter space of the simplified rupture model where the two non-dimensional parameters are $\bar \tau$ and $\bar \gamma$. The transition from crack-like to pulse-like propagation is governed by the shear pre-stress distribution at the onset of rupture (through the parameter $\bar \tau$), and whether there is a stress boundary condition or a no-slip boundary condition at the top interface (through the parameter $\Gamma$).
\label{fig:sketch}}
\end{figure}

\section{The crack-pulse transition \label{sec:crack-pulse}}
We initialize $\bar \tau (\bar x)$ with a (small) Gaussian function with maximum $\bar \tau = 1$ and standard deviation 1, so that rupture nucleation occurs at $\bar x = 0$ (\figurename~\ref{fig:sketch}). We use either a symmetric or an asymmetric pre-stress. If the pre-stress is symmetric, it reaches a constant value $\bar \tau_{0,r} = \bar \tau_{0,l} = \bar \tau_0$ in both directions, where subscripts $r,l$ refer to right and left. If pre-stress is asymmetric, it reaches a positive value $\bar \tau_{0,r} \geq 0$ for increasing $\bar x$ and a negative value $\bar \tau_{0,l}$ for decreasing $\bar x$ (see the insets of \figurename~\ref{fig:sketch} for a visual representation of these definitions).

For a symmetric pre-stress distribution with $\bar \tau_0 > 0$, we observe bi-directional propagation of either a crack (stress boundary condition with $\Gamma = 0$) or a crack that transitions into two slip pulses (displacement boundary condition, $\Gamma > 0$). For asymmetric pre-stress with $\bar \tau_{0,+} > 0$ in one direction and $\bar \tau_{0,-} < 0 $ in the other direction, we observe cracks that transition to unidirectional slip pulses for both $\Gamma = 0$ and for $\Gamma > 0$ if $\bar \tau_{0,r} = 0$ in one direction and $\bar \tau_{0,l} < 0 $ in the other direction. This model behavior is summarized in \figurename~\ref{fig:sketch}.

\subsection{Steady-state slip pulse}
\begin{figure}
\includegraphics[]{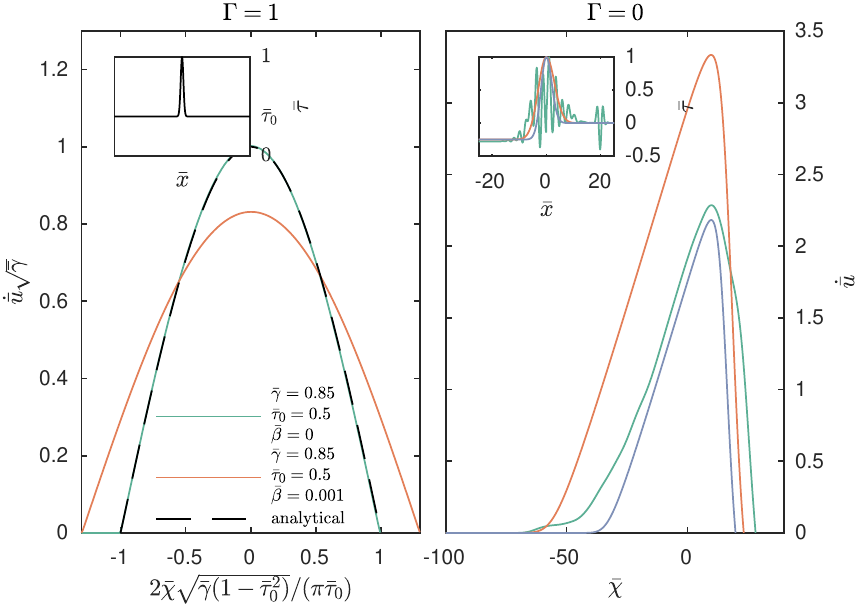}
\caption{\emph{Left}: numerically derived steady-state solution (in the co-moving frame $\bar \chi$) of the slip velocity inside the pulse under constant displacement boundary condition, $\Gamma=1$, compared to the analytical solution. The shear pre-stress profile is given in the inset, where a Gaussian function with standard deviation 1 in the center of the domain is used to nucleate rupture. The figure also shows the solution for non-zero $\bar \beta$. \emph{Right}: numerically-derived steady-state slip speed of slip pulses under constant stress boundary condition, $\Gamma = 0$. Pulses remain in steady-state because they propagate in a region of $\bar \tau_{0,r} = 0$. The corresponding shear pre-stress $\bar \tau (\bar x)$ is shown in the inset. In this case, the slip velocity of the pulse is entirely determined by $\bar \tau (\bar x)$ at nucleation.
\label{fig:steady_state}}
\end{figure}
For $\Gamma > 0$ and $\bar \tau > 0$, the steady-state solution of dynamic frictional rupture is pulse-like. In the limit of $\bar \beta = 0$, this steady-state solution for a slip pulse propagating in a region of constant pre-stress $\bar \tau_0$ can be calculated in closed form and the derivation is provided in the appendix \ref{app:analytical}. The slip velocity is a cosine
\begin{align}
\dot{\bar u} = \frac{1}{\sqrt{\bar \gamma}}
\cos \left ( \frac{\sqrt{\bar \gamma (1-\bar \tau_0^2)}}{\bar \tau_0} \bar \chi \right ),
\end{align}
in the co-moving frame where $\bar \chi = 0$ corresponds to the position of the maximum slip velocity, which is given by $\dot{\bar u}_\text{max} = \frac{1}{\sqrt{\bar \gamma}}$, and the pulse width is $\bar W = \frac{\pi \bar \tau_0}{\sqrt{\bar \gamma (1-\bar \tau_0^2)} }$. To obtain this solution we also used the steady-state front velocity $\bar v_c = \frac{1}{\sqrt{1-\bar \tau_0^2}}$ which has been obtained previously by \citet{amundsen2015steady}.
\figurename~\ref{fig:steady_state} shows the steady-state slip speed inside a pulse for $\Gamma = 1 $ as well as $\Gamma = 0$, which correspond to displacement and stress boundary conditions applied on the top surface, respectively. For $\Gamma = 1$, the pulse shape is well defined. For $\Gamma = 0$, a unique solution of a steady-state pulse does not exist because under uniform stress $\bar L_\text{trans}$ and $\bar T_\text{trans}$ tend to infinity when $\bar \gamma \rightarrow 0$. However, under non-uniform pre-stress conditions, pulses can still nucleate for $\Gamma = 0$. In this case, the pulse size will be entirely determined by the rupture history which is controlled by the initial pre-stress distribution, and is only stable when propagating in a region of $\bar \tau = 0$, i.e. the pulse grows when $\bar \tau > 0$ and decays when $\bar \tau < 0$.

\begin{figure}
\includegraphics[]{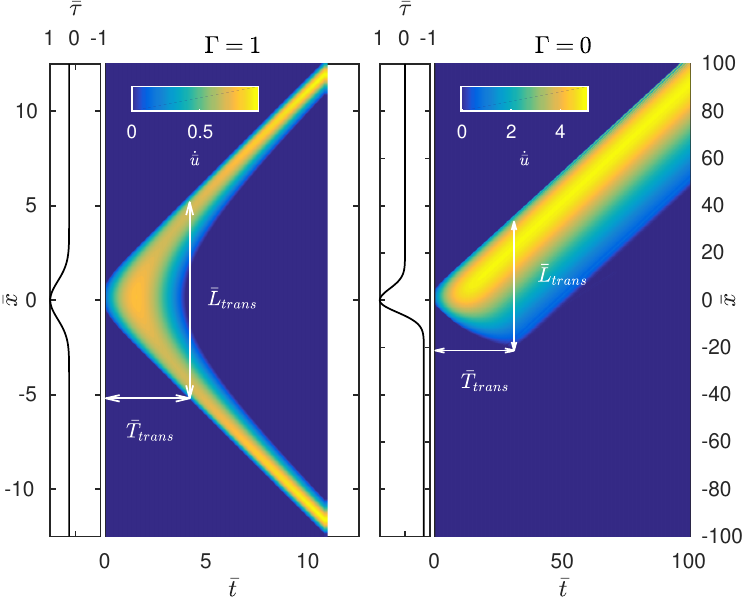}
\caption{\emph{Left}: Spatio-temporal slip velocity using $\Gamma = 1$ and a symmetric stress. A (small) Gaussian function with standard deviation 1 is applied in the center of the domain to nucleate rupture. \emph{Right}: Spatio-temporal slip velocity for $\Gamma = 0$ with a non-symmetric pre-stress distribution (left graph). The figure contains the definitions of the transition time $\bar T_\text{trans}$ and the transition length $\bar L_\text{trans}$.
\label{fig:spatiotemporal}}
\end{figure}

\subsection{Crack-pulse transition for slip boundary condition (\texorpdfstring{$\Gamma = 1$}{})}
For displacement boundary condition applied at the top surface, we can estimate the transition from crack to pulse using the analytical steady-state solution for the slip pulse. The transition time $\bar T_\text{trans}$ is defined by the time it takes to reach a slip of $\bar u_\text{trans} = \frac{2\bar \tau}{\bar \gamma}$ at the point of nucleation. If we assume that $\frac{\partial^2 \bar u}{\partial \bar x^2}$ is small, the equation of motion for the nucleation point can be approximated as $\ddot {\bar u} (\bar t) \approx -\bar \gamma \bar u (\bar t) + \bar \tau$, which under the initial conditions $\bar u(0) = 0$ and $\dot {\bar u}(0) = 0$ has the solution $\bar u(\bar t) = \frac{\bar \tau}{\bar \gamma} \left (1 - \cos (\sqrt{\bar \gamma} \bar t) \right )$. $\bar T_\text{trans}$ is found from $\bar u(\bar T_\text{trans}) = 2 \frac{\bar \tau}{\bar \gamma}$ where we use the first solution $\bar T_\text{trans} = \frac{\pi}{\sqrt{\bar \gamma}}$. Assuming a uniform pre-stress $\bar \tau(\bar x) = \bar \tau_0$ and using $\bar T_\text{trans}$, we find $\bar L_\text{trans}$ using the front speed $\bar v_c$:
\begin{equation}
    \bar L_\text{trans} = 2 \bar T_\text{trans} \bar v_c = \frac{2 \pi}{\sqrt{\bar \gamma(1-\bar \tau_0^2)}}.
\end{equation}
\figurename~\ref{fig:spatiotemporal} (left) shows an example of a crack-pulse transition at $\Gamma = 1$, in agreement with the predictions derived in this section.

\subsection{Crack-pulse transition for stress boundary condition (\texorpdfstring{$\Gamma = 0$}{})}
For stress boundary condition ($\Gamma = 0$) pulses are inherently unstable, and the pulse shape is not unique. Pulses only keep a steady shape if propagating in a region where the available elastic strain energy exactly corresponds to the dissipation in the pulse, which for Amontons-Coulomb friction corresponds to $\bar \tau = 0$. This can be seen from equation \ref{eq:eom} which reduces to the one-dimensional wave equation when $\bar \tau = 0$ and $\Gamma = 0$. Any small perturbation in $\bar \tau$ can cause the pulse to change its shape, vanish or transition to a crack. For more complicated friction laws or when solving the system in more than one dimension, it will in general be very unlikely to nucleate a pulse where dissipation exactly matches the available elastic strain energy.
For $\Gamma = 0$, the onset of a steady-state pulse requires that the following two conditions are met together. i) The pre-stress is such that an initial crack arrests in one direction and not the other. This requires an asymmetry in the pre-stress. ii) The pre-stress in the propagating direction reaches zero at some point. For $\Gamma = 0$, $L_\text{trans}$ is determined by the crack size when the crack arrests in one direction. The position of front arrest can be approximated by noting that stress balance at zero acceleration requires that $\bar \tau = 0$ in the ruptured region. Assuming nucleation of rupture is located at $\bar x = 0$ and the arresting occurs in the negative direction, we can write 
\begin{equation}
\int_{x_\text{-}}^\infty \bar \tau(x) = 0
\label{eq:x_minus}
\end{equation}
which can be used to determine $\bar x_\text{-}$ for any $\bar \tau (\bar x)$. The transition time is given by the time it takes to propagate from $\bar x = 0$ to $\bar x_\text{-}$, which can be found if the rupture speed $\bar v_c$ is known.
\begin{equation}
\bar T_\text{trans} = \frac{|\bar x_\text{-}|}{ \langle \bar v_c \rangle_\text{-}}
\end{equation}
The total crack size is
\begin{equation}
\bar L_\text{trans} = |\bar x_\text{-}| + |\bar x_\text{+}|
\end{equation}
where
\begin{equation}
\bar x_\text{+} = \int_0^{\bar T_\text{trans}} \bar v_{c,+} \text{d} \bar t
\end{equation}
In the arresting direction, $\bar \tau$ has to be negative. Then, we do not have an analytical expression for the front speed, but we instead use $\langle \bar v_c\rangle_\pm \approx 1$ leading to $\bar T_\text{trans} \sim |\bar x_-|$. and $\bar L_\text{trans} \sim 2|\bar x_-|$, where $\bar x_-$ is found from equation \ref{eq:x_minus}. Note that this will lead to a slight underestimation of $\bar T_\text{trans}$. \figurename~\ref{fig:spatiotemporal} (right) shows an example of a crack-pulse transition at $\bar \gamma = 0$, in agreement with the predictions derived in this section.

\begin{figure}
\includegraphics[]{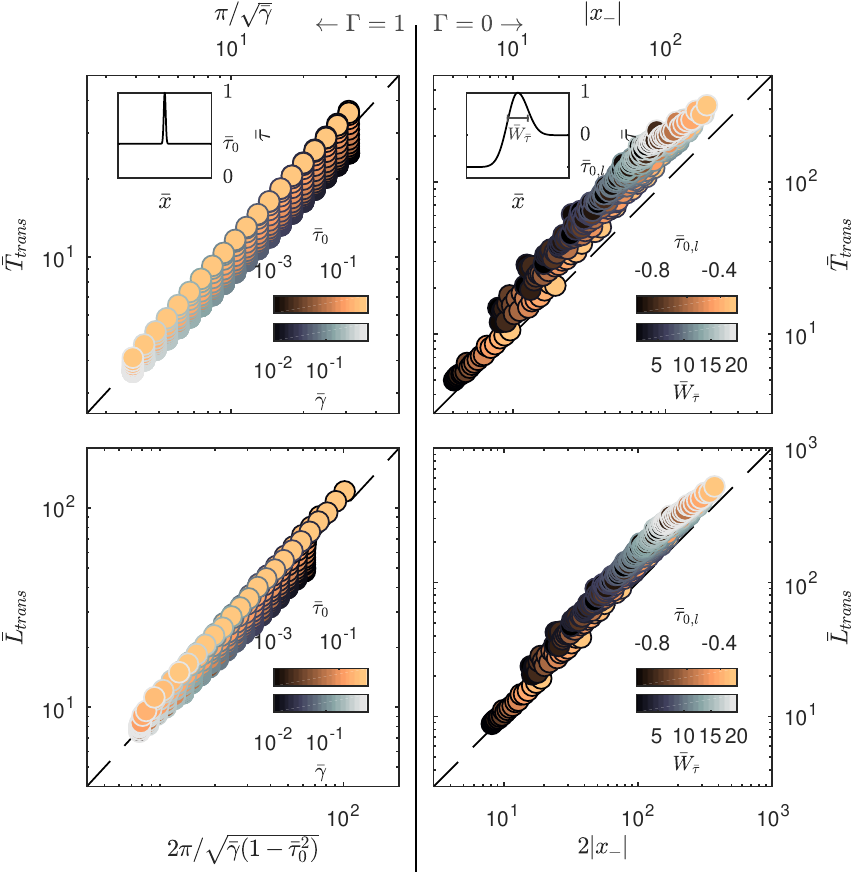}
\caption{Scaling relations of $\bar L_\text{trans}$ and $\bar T_\text{trans}$ for the crack-pulse transition for both $\Gamma = 1$ (left panels), and $\Gamma = 0$ (right panels). For $\Gamma =0$, we use an asymmetric stress distribution (top right inset), while for $\Gamma = 1$ we use a symmetric stress distribution (top left inset). The face color shows the stress $\bar \tau_0$ or $\bar \tau_{0,l}$, while the color of the rim shows $\bar \gamma$ or $\bar W_{\bar \tau}$.
\label{fig:scaling}}
\end{figure}
We have calculated the scaling relations for $\Gamma = 1$ and $\Gamma = 0$ over a wide range of $\bar \gamma$ and $\bar \tau$. While the non-dimensional elasticity ratio $\bar \gamma$ is expected to be found in a limited range close to $2/3$ \cite{ji2010lame}, for completeness we include a wider range. For $\Gamma = 0$, we systematically vary the initial width $\bar W_{\bar \tau}$ of the pre-stress perturbation to obtain different transition lengths and times. \figurename~\ref{fig:scaling} shows the data collapse using the scaling relations derived above. The model and analytical results are in good agreement for the prediction of the transition from crack-like to pulse-like propagation.

\section{Discussion \label{sec:discussion}}
In this discussion, we first insert dimensional quantities in the non-dimensional scaling relations found above, and qualitatively address how these predictions relate to existing literature. We then move on to put this work in the context of previously discussed mechanisms for the onset of slip pulses along frictional interfaces. We then discuss stability of pulses, and the role boundary conditions in pulse stability. 
\subsection{Predicted scaling relations from the minimal model}
In our model, the onset of slip-pulse frictional rupture occurs through a transition from crack-like to pulse-like rupture at a crack size that depends on pre-stress and boundary conditions. For stress boundary conditions, the transition length is controlled in full by the initial pre-stress distribution, and the transition occurs because the crack arrests in one direction and not the other. The pre-stress in the propagating direction is set so that dissipation in the pulse is equal to the available elastic strain energy. For slip boundary conditions, the transition from crack to a self-similar pulse is given by the system width $H$, the elasticity ratio $\bar \gamma$ as well as the ratio of rupture speed $v_c$ and speed of sound $v_s$.

When inserting dimensional quantities we obtain $L_\text{trans} = 2\pi H\sqrt{\frac{\lambda + 2G}{2G}} \frac{v_c}{v_s}$. The pulse width is $W = \pi \bar \tau H \sqrt{\frac{\lambda + 2G}{2G}} \frac{v_c}{v_s}$ with a maximum velocity $\dot u_\text{max} = \frac{\Delta \tau}{\sqrt{2G \rho}}$, where $\Delta \tau = (\mu_s - \mu_k)\sigma_n$ is the stress drop. The model predicts that the onset of slip pulses occurs at shorter propagation lengths in systems with small $H$, and low initial pre-stress. Applied to earthquake mechanics, these results are consistent with i) the observation that large earthquakes favor slip-pulses \cite{melgar2017systematic}, ii) the observation that large aspect ratios seem to favor the formation of slip pulses \cite{nielsen2000influence}, iii) the observation that increasing pre-stress widens slip-pulses so that they approach crack-like solutions \cite{Lu2010pulse}, and iv) the prediction that the seismogenic depth may limit the width of slip pulses \cite{ampuero2017upper,bai2017effect}.

\subsection{Relation to other mechanisms for the onset of slip pulses}
The present study introduces a one-dimensional two-parameter continuum model for the onset of slip-pulses which can be understood from elastic relaxation and redistribution of elastic energy stored in the pre-stress only. In that respect, it is useful to place this minimal model in the context of previously discussed mechanisms for the onset of slip pulses along frictional interfaces. Three main mechanisms are often encountered in the literature: i) velocity-weakening friction, ii) coupling between slip and normal stress, iii) stress barriers and heterogeneous stress distributions. Below we discuss the relation between this study and these three mechanisms.

\subsubsection{Velocity-weakening friction}
A widespread explanation for the onset of slip pulses is velocity-weakening dynamic friction. This idea was first introduced by \citet{heaton1990evidence}. The argument is as follows. First, one assumes that a ruptured part of a frictional interface can be described using dislocation theory. Slip is then expected to vary as $u \sim \sqrt{v_c t - x}$ in the region behind the rupture front, with slip speed $\dot u$ decreasing as the inverse square root of $x$ behind the front. Assuming linear velocity weakening friction ($f_f = f_s - \alpha \dot u$), the frictional resistance is then expected to increase with increasing distance from the rupture front. In turn, this causes the crack to heal, causing the propagation of a self-healing slip pulse. Velocity-weakening friction has since been identified as an important ingredient for the existence of slip pulses. In particular, several studies have reported a transition between crack-like and pulse like solutions as the characteristic velocity that controls the velocity-weakening steady state is increased \cite{nielsen2000influence,nielsen2000rupture,cochard1994dynamic,zheng1998conditions}.

\subsubsection{Coupling between slip and normal stress}
A different but related mechanism that has been previously identified to generate slip-pulses, is weakening of the frictional resistance within slip pulses due to a coupled response between slip along the interface and normal stress, even to the point of surface separation \cite{Schallamach71waves,gerde2001friction}. This can occur in a number of ways. First, a material contrast across the frictional interface will cause such coupling. Slip along such a bi-material interface will cause a break of stress symmetry. In turn, this leads to the prediction that local normal stress is directly coupled to interface slip \cite{shlomai2016structure,andrews1997wrinkle}. Experiments have demonstrated that pulses then propagate in the direction of motion of the soft material, while the opposite direction favors crack-like propagation \cite{shlomai2016structure}. Second, the existence of pore fluids along the interface can cause such coupling. In this case, slip can directly couple to normal stress through elevated pore pressures due to thermal pressurization. In this case, slip-pulses can arise in the absence of velocity-weakening friction, material contrasts and heterogeneous stress \cite{garagash2012seismic}.

\subsubsection{Stress barriers and heterogeneous stress distributions}
A third mechanism that has been proposed is the existence of stress barriers and heterogeneous stress distributions along frictional interfaces. \citet{johnson1990initiation} demonstrated that frictional rupture starting out as crack-like can transition to pulse-like if it arrests in one of the propagating directions. Subsequently, a healing front is initiated at the barrier, propagating backwards causing a uni-directional pulse. This mechanism can be extended to heterogeneous stress distributions \cite{beroza1996short,day1998dynamic}. In such case, the heterogeneous stress distribution can be seen as multiple stress barriers causing a large variety of propagating and healing fronts leading to a transition from crack-like propagation to uni-directional pulse-like propagation.

\subsubsection{Mechanism of slip pulses in the minimal model}
The slip pulses in the minimal model are not a consequence of velocity-weakening dynamic friction. In principle, one could argue that Amontons-Coulomb friction is both velocity weakening and slip weakening where the weakening occurs over an infinitesimal velocity or displacement step. However, the argument carried out by \citet{heaton1990evidence} does not apply to this case as it requires dynamic friction to be velocity weakening. 

The slip pulses in the minimal model do not rely on a dependency of normal stress on slip. While our setup is bi-material (elastic-rigid), the bi-material coupling is not present in our simulation because of the constraint set on the vertical displacements $\dot u_y = 0$. This assumption ensures there is no feedback between slip along the frictional interface $u_y$ and the normal stress at the interface $\sigma_n$.

For stress boundary conditions, a stress barrier is necessary in the minimal model to transition from crack-like to pulse-like rupture.  Further, a steady state solution only exists if the pulse can propagate along an interface where the energy available in the form of pre-stress exactly balances the energy dissipated by the pulse. In other words, the minimal model mechanism of slip pulses for stress boundary conditions is the same as the above mentioned mechanism of stress barriers.

For displacement boundary conditions the consequence of a stress barrier in the minimal model would be a change from bi-directional to uni-directional pulse-propagation. However, the barrier itself is not a necessary condition for the emergence of pulse-like propagation. In the minimal model, slip pulses under displacement boundary condition at the top interface, occur because the resistance to slip is increased with slip because of elastic relaxation and a finite thickness. This mechanisms does not depend on details of the friction law, and slip pulses are thus expected to be generic features of systems with imposed displacement boundary condition. In that sense, our model offers a $4^{th}$ basic mechanism for slip-pulse creation that arises simply from elastic interactions with the imposed boundary condition.

\subsubsection*{Mechanism of slip pulses in the minimal model, in context of the discrete Burridge-Knopoff model}
\begin{figure*}
    \centering
    \includegraphics[width=\textwidth]{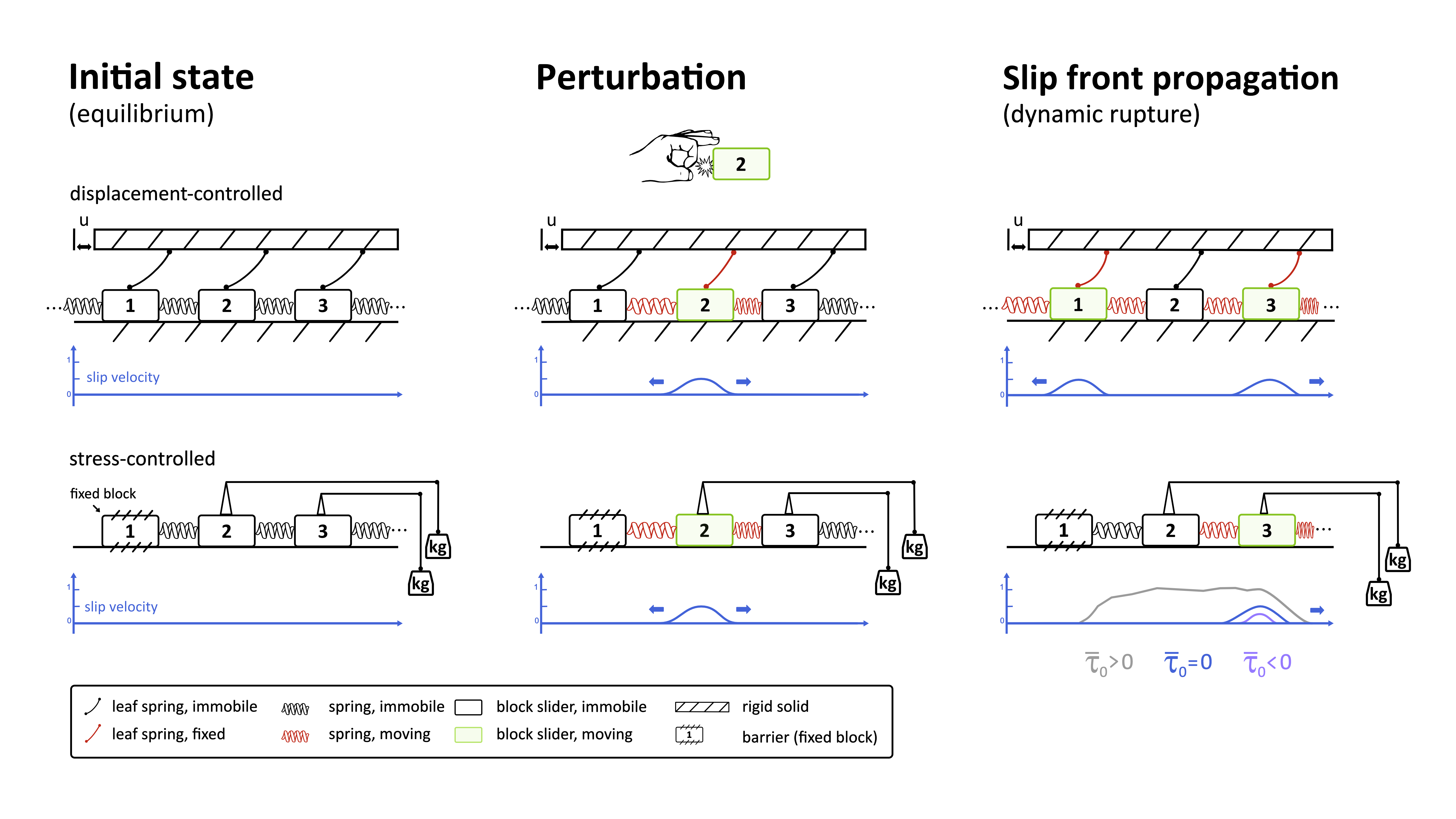}
    \caption{Sketch of onset of slip pulses in the discrete Burridge-Knopoff model. The Burridge-Knopoff model is a special case of the finite difference discretization of the continuum model used here. The top row shows a discrete version of \figurename~\ref{fig:sketch} with $\Gamma = 1$, with corresponding onset of slip pulses in the continuous limit shown in \figurename~\ref{fig:spatiotemporal} (left). The bottom row shows a discrete version of \figurename~\ref{fig:sketch} with $\Gamma = 0$ with corresponding onset of slip pulses in the continuous limit shown in \figurename~\ref{fig:spatiotemporal} (right).}
    \label{fig:pulse_BK_sketch}
\end{figure*}
The minimal model used here is directly connected to the Burridge-Knopoff model (see e.g. \citet{amundsen2015steady} and \cite{carlson1989mechanical}) and our equation~\ref{eq:eom}. The Burridge-Knopoff model is a special case of the finite difference discretization of equation~\ref{eq:eom} when $dx = H$. Because the Burridge-Knopoff model is a widely used model in the study of frictional systems,  especially in the context of earthquake physics, we dedicate this section to an explanation of the physical mechanism of our findings in context of the discrete Burridge-Knopoff model.

The Burridge-Knopoff model is illustrated in \figurename~\ref{fig:pulse_BK_sketch}. Our continuum model in equation~\ref{eq:eom}, in its discrete form, essentially describes a mass connected by Hookean identical springs to its neighbors, sitting on a frictional surface and driven by either a stress or strain boundary condition from \emph{above}. The $ \gamma \bar u$ term represents a leaf-spring that connects each block to the driving plate above, where $\Gamma$ determines if the leaf springs are present or not. The $\bar \tau$ term represents the space-varying initially stored elastic stress in excess of the frictional resistance. The $\beta$ term provides a viscous damping term that reduces high frequency oscillations. 

\figurename~\ref{fig:pulse_BK_sketch} top illustrates constant displacement boundary condition imposed at the top wall. This drives masses to move via forces imparted by leaf springs.  If a single block starts sliding to the right the driving force from its leaf spring drops, yet the block displacement has stretched its neighboring springs. Now the 2 neighboring blocks may start to slide due to the sum of the drive from their leaf spring and their normal springs. The motion of the blocks relaxes their leaf spring forces which slows them down, while loading adjacent springs and driving their neighbors to move. The figure shows single block wide pulses moving right and left.

A constant stress boundary condition in equation~\ref{eq:eom} ($\Gamma=0$) implies no leaf springs, and that the system is driven by stress applied on each block. This is mathematically equivalent an initial relative displacement of all blocks \cite{amundsen2015steady,thogersen2019minimal}. A slip pulse in \figurename~\ref{fig:pulse_BK_sketch} lower panels will initiate when one of the blocks is initially given a larger potential energy than its neighbors (a spike in $\bar \tau$). Such a perturbation in elastic energy will travel as a self-similar wave solution if the following conditions are met: i) There is a barrier leading to rupture arrest in one of the directions (stuck block in lower panels of \figurename~\ref{fig:pulse_BK_sketch}). ii) The pre-stress along the direction of propagation reaches the dynamic friction level or below. The pulse will grow or decay depending on whether the available elastic energy, in the form of pre-stress it encounters on its way, is $\bar \tau_0 > 0 $ or $\bar \tau_0 < 0$. A steady-state solution only exists if the pulse propagates in a region with $\bar \tau_0 = 0$.

\subsection{Stability of slip pulses -- the crucial role of boundary conditions}

\subsubsection{Pulse stability -- stress versus displacement boundary conditions}
Recent studies have demonstrated the inherent instability of slip-pulses for different kinds of friction constitutive laws \cite{gabriel2012transition,brener2018unstable,noda2009earthquake,brantut2019stability}. These models report the existence of a steady-state pulse solution located at the sharp transition between growing pulses, whose spatial extent increases during propagation, and decaying pulses, whose spatial extent progressively shrinks and eventually arrests the pulse. The slip pulses in our simplified model have the same intrinsic instability only when stress boundary conditions are used ($\Gamma = 0$). Then, slip pulses only keep their shape when the available elastic strain energy in the form of pre-stress is exactly equal to the dissipation in the pulse. For our choice of Amontons-Coulomb friction law, this corresponds to $\bar \tau = 0$, when the equation of motion reduces to the one-dimensional wave equation. In this case, the net stress change is zero if the pulse velocity profile is constant, and the net stress change is non-zero only if the pulse accelerates or decelerates. This is consistent with the observation that pulses can propagate rapidly with negligible net stress change \cite{mclaskey2015slip}. Any small perturbation in $\bar \tau$ will make the slip pulse decrease or grow. The velocity profile of these unstable pulses under constant stress boundary conditions ($\Gamma = 0$) are not unique in our model. This non-uniqueness is likely a feature of our specific choice of friction law. For different friction laws, unique solutions may be likely found through couplings between sliding velocity, slip and pulse dissipation. Interestingly, a slip boundary condition stabilizes the slip pulses through elastic relaxation. Steady-state slip pulses are then found for any pre-stress $\bar \tau \in (0,1)$, and perturbations in $\bar \tau$ will simply transition pulses from one steady-state solution to another.

\subsubsection{The characteristic length $H$ in the context of earthquake dynamics}
In the context of fault mechanics, $H$ can naturally be interpreted as the distance from the fault at which tectonic displacements are imposed. For the example of subduction zones, $H$ can then be understood as the crustal thickness. Nevertheless, to enable the slip pulse mechanism proposed in this paper (i.e. elastic relaxation from imposed displacement at the boundary), the duration of the rupture $t^*$ must be sufficiently large to allow elastic waves (typically moving at the shear wave speed $c_s$) to travel back and forth from the plate boundaries: $t^*>2H/c_s$. In light of this criterion, the proposed model provides an explanation why slip pulse becomes the dominant rupture mode for large earthquakes \cite{melgar2017systematic} but also slow-slip events \cite{dalzilio2020slowslip} which are both characterized by long rupture duration. 

\begin{figure}
\includegraphics[width=0.75\linewidth]{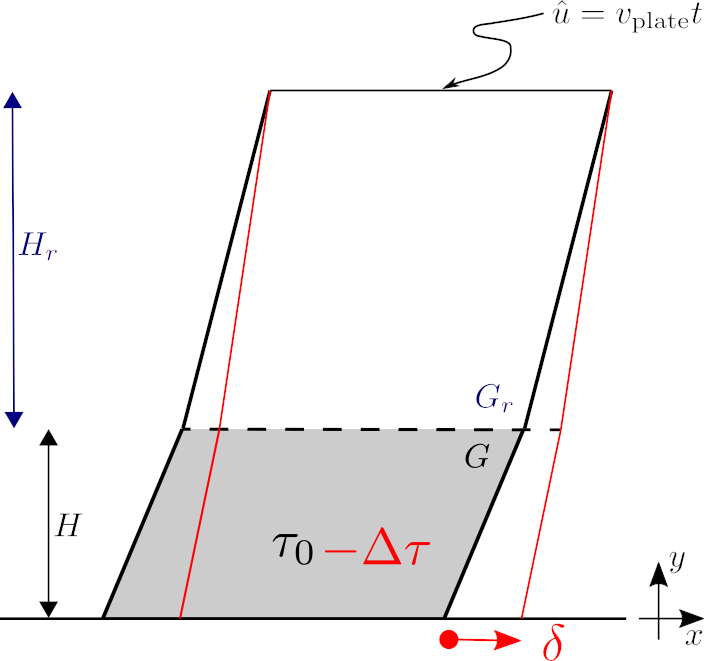}
\caption{Schematic deformation of a fault zone with a damaged core. A symmetric configuration can be assumed in the $y<0$ direction with no loss of generality. \label{fig:damage_zone}}
\end{figure}

Along mature fault zones, the interface between the stiff host rock and the more compliant damage zone can reflect part of the radiated waves, enhancing self-healing in the wake of a rupture and easing the nucleation of slip pulses \cite{huang2011pulselike,huang2014earthquake,thakur2020effects}. Recently, \citet{idini2020faultzone} showed how fault core damage zone also changes quasi-static stress relaxation and favours pulse-like rupture style. In light of our model, such effect corresponds to a reduction in the characteristic length $H$, which now scales as the thickness of the damage zone. This can be understood from the idealized 2D geometry of a mature fault with damage zone presented in \figurename~\ref{fig:damage_zone}. The fault core damage zone is surrounded by a host rock of different shear modulus $G_r>G$. The system is sheared by a far-field displacement $\hat u$ applied at a distance $H+H_r$ from the fault. As the plate velocity loading the system is very small compared to the slip velocity, we can assume that $\hat u$ is a far-field imposed displacement over the duration of the rupture and analyze the static stress drop after the interface has slipped a distance $\delta$.
From the continuity of stress and displacement at $y=H$, the stress drop $\Delta\tau$ associated to interfacial slip $\delta$ writes:
\begin{equation}
 \Delta\tau = G\Delta\gamma = G_r\Delta\gamma_r,
\end{equation}
with $\Delta\gamma_r$ and $\Delta\gamma$ being respectively the change of shear strain in the rock and the damage zone, such that
\begin{equation}
 \delta = H\Delta\gamma + H_r\Delta\gamma_r = \Delta\gamma\left(H + \frac{G}{G_r}H_r \right).
\end{equation}
Combining the two equations above, one gets that
\begin{equation}
 \Delta\tau = G\frac{\delta}{H + \frac{G}{G_r}H_r}.
 \label{drop}
\end{equation}
In the limit of a high stiffness contrast ($G\ll G_r$), Eq. (\ref{drop}) reduces to $\Delta\tau=G\delta/H$, which is exactly equivalent to the stress drop obtained if the imposed displacement $\hat u$ is directly set at the boundary $y=H$ (rigid motion of the wall rock).
In this context, our study explains how the formation of a compliant fault core damage zone plays a role on the existence and the stability of steady-state pulse solutions. Whereas the instability of slip pulses has been demonstrated for homogeneously imposed stress conditions and diverse friction models \cite{gabriel2012transition,brener2018unstable,noda2009earthquake,brantut2019stability}, the development of off-fault damage can turn the system into imposed slip boundary conditions, for which stable steady-state pulse solutions exist, even for Amontons-Coulomb friction. The existence of generic pulse solutions suggests that slip pulses could be the dominant mode of frictional rupture along highly damaged fault zones.

\section{Concluding remarks \label{sec:concusion}}
In view of the number of features of slip pulses captured by the minimal model presented here, we propose that pulse-like propagation along frictional interfaces is a likely generic feature. While different friction laws will yield different velocity profiles, crack-pulse transition lengths, propagation speeds and pulse widths, the existence of slip pulses does not rely on details in the friction constitutive law in our model. Instead, the pre-stress and the boundary conditions alone can produce slip pulse solutions along frictional interfaces. This interpretation has important consequences. In particular, in laboratory experiments and natural earthquakes, one should be careful with interpreting observations of slip pulses as proof of velocity-weakening friction.

Preventing the appearance of slip pulses is in many situations desirable as they are the origin of squealing noise in train breaks and several industrial processes \cite{kinkaid2003automotive,chen2019effect}. Our findings also demonstrate the prospect of controlling the onset of slip pulses as well as their stability through i) manipulation of the pre-stress at the frictional interface and ii) control of the system boundary conditions. We can sum up our findings in three main points:
\begin{enumerate}
    \item We have shown that slip pulses occur in a minimal continuum friction model with only two non-dimensional parameters. The transition from crack-like to pulse-like propagation as well as the pulse width scales with the system thickness, the rupture speed, the pre-stress and the elastic parameters of the material for displacement boundary conditions. For stress boundary conditions, the transition and the pulse width are entirely determined by the non-dimensional pre-stress.
    \item The onset of slip pulses can be explained by redistribution of elastic stress alone, which means that slip pulses are likely generic phenomena that exist for a wide range of friction constitutive laws.
    \item The stability of slip pulses is largely influenced by the system boundary conditions. In our minimal model, stress boundary conditions favor unstable slip pulses, while displacement boundary conditions favor stable slip pulses.
\end{enumerate}

\acknowledgements{K.T. acknowledges support from the Norwegian Research Council through FRIPRO (Grant 287084). E.A thanks ISF grant 910/17. F.B. acknowledges support of the Swiss National Science Foundation through the fellowship No. P2ELP2/188034. We thank Pablo Ampuero for an interesting discussion and valuable feedback on the manuscript.}

\bibliographystyle{apsrev4-1}
\bibliography{main.bbl}

\appendix

\section{Equation of motion \label{app:eom}}
Here, we supplement the main text with the derivation of the equation of motion. We start with the integration over the red rectangle in \figurename~\ref{fig:BC_sketch}:
\begin{widetext}
\begin{align}
    H \rho \langle \frac{\partial^2 \vec u}{\partial t^2} \rangle_y = \int_0^H \nabla \cdot \boldsymbol{\sigma} \text{d}y & =  \frac{\partial }{\partial x} \int_0^{\text{d}x} \int_0^H \nabla \cdot \boldsymbol{\sigma} \text{d}x \text{d}y =  \frac{\partial}{\partial x} \oint_S \boldsymbol{\sigma} \vec n \text{d}S \\
    & = \frac{\partial }{\partial x}\left ( \text{d}x \boldsymbol{\sigma} \vec n |_{y=H} + \text{d}x \boldsymbol{\sigma} \vec n |_{y=0} + H \langle \boldsymbol{\sigma}\vec n|_{x=\text{d}x}\rangle_y + H \langle \boldsymbol{\sigma}\vec n|_{x=0}\rangle_y \right )\\
    & = \boldsymbol{\sigma} \vec n |_{y=H} + \boldsymbol{\sigma} \vec n |_{y=0} + \frac{\partial}{\partial x} H [\langle \sigma_{xx}\rangle_y, \langle \sigma_{xy} \rangle_y]^T 
\end{align}
\end{widetext}
Here, $S$ denotes the surface of the volume over which the integration is calculated, and $\vec n$ is the normal vector. The boundary conditions can now be set through the surface tractions $[\hat \sigma_x(H),\hat \sigma_y(H)]^T \equiv \boldsymbol{\sigma} \vec n |_{y=H}$ and $[\hat \sigma_x(0),\hat \sigma_y(0)]^T \equiv \boldsymbol{\sigma} \vec n |_{y=0}$. We further assume $u_y = 0$ which results in the last term being reduced to $\frac{\partial}{\partial x} H [\sigma_{xx}, 0]$, which is the assumption needed to be able to consider only the x-coordinate. At the frictional interface, $\hat \sigma_x(x,0,t)$ is set by the friction law. At the top interface, we either set a constant traction, or a constant displacement (i.e. no slip boundary).

Now we insert for $u_x(x,t)$ and assume $u_y \approx 0$ with constant $\lambda$ and $G$, and we obtain
\begin{equation}
H \rho \frac{\partial^2\langle u_x \rangle_y}{\partial t^2} = H  \frac{\partial \langle \sigma_{xx} \rangle_y}{\partial x} + \hat \sigma_{x}(x,H,t) - \hat \sigma_x(x,0,t),
\end{equation}
where we have introduced a minus sign on $\hat \sigma_x(x,0,t)$ in order to consider the $f_f$ as positive in equation \ref{eq:ff_app}. Inserting for $\sigma_{xx}$, we obtain
\begin{equation}
H \rho \frac{\partial^2 \langle u_x \rangle_y}{\partial t^2} = H (\lambda + 2G)\frac{\partial^2 \langle u_x \rangle_y}{\partial x^2} +  \hat \sigma_{x}(x,H,t) - \hat \sigma_x(x,0,t).
\end{equation}
The friction law at the interface can be applied through $\hat \sigma_{x}(0,t)$, and $\hat \sigma_{x}(H,t)$ is the driving stress. $y=0$ defines the frictional interface location. We use two variations of the boundary condition $\hat \sigma_{x}(x,H,t)$.

First, we follow what was used by \citet{bouchbinder2011slow} and apply a shear stress boundary condition. $\hat \sigma_{x}(x,H,t)$ is then set as a constant function $\tau(x)$ that does not vary in time. The equation of motion can then be written as
\begin{equation}
H \rho \frac{\partial^2 \langle u_x \rangle_y}{\partial t^2} = H (\lambda + 2G)\frac{\partial^2 \langle u_x \rangle_y}{\partial x^2} + \tau - \hat \sigma_{x}(x,0,t).
\end{equation}
To be able to apply a friction law at $y=0$, we need to make an additional approximation that relates the average slip with the slip at the interface. When we later will apply Amontons-Coulomb friction at $y=0$, it is sufficient to assume $\text{sign}(\dot u(y=0)) = \text{sign}(\langle \dot u \rangle)$. Alternatively, for velocity-dependent friction laws an alternative approach would be to approximate $\dot u(y=0) \approx \langle \dot u \rangle$.

Second, it is also possible to introduce a constant displacement (i.e. no slip) boundary condition at $y=H$; $\frac{\partial u_x (H)}{\partial t} = 0$. This can be rewritten in terms of a driving stress which is set by a relative displacement at the top boundary and the frictional interface $u_x(x,H,t)-u_x(x,0,t)$. Then, we need to introduce an approximate relation between the displacement at $y=H$ and the displacement at $y=0$, which we do by linearizing the displacement profile. The shear stress $\hat \sigma_{x}(x,H,t)$ can then be rewritten in terms of the imposed displacement at the top boundary, matching the traction and the shear stress in the volume close to the interface $G \frac{\partial u_x}{\partial y}|_{y=H}$.
\begin{widetext}
\begin{equation}
H \rho \frac{\partial^2 \langle u_x \rangle_y}{\partial t^2} = H (\lambda + 2G)\frac{\partial^2 \langle u_x \rangle_y}{\partial x^2} + \frac{u_x(x,H,0) - u_x(x,0,t)}{H} G - \hat \sigma_{x}(x,0,t).
\end{equation}
\end{widetext}
To evaluate the relative slip of the frictional interface and the top layer we can no longer use the averaged value $\langle u_x \rangle_y$, but instead change the slip of interest to $u_x(x,0,t)$.
\begin{widetext}
\begin{equation}
\frac{1}{2}H \rho \frac{\partial^2 u_x(x,0,t)}{\partial t^2} = H \frac{\lambda + 2G}{2}\frac{\partial^2 u_x(x,0,t)}{\partial x^2} - H \frac{\lambda + 2G}{2}\frac{\partial^2 u_x(x,H,0)}{\partial x^2} + \frac{u_x(x,H,0) - u_x(x,0,t)}{H} G - \hat \sigma_{x}(x,0,t).
\end{equation}
\end{widetext}
where $\frac{\partial^2 u_x(x,H,0)}{\partial x^2}$ is a constant function that arises if the the displacement at the top interface is not set as constant.
At $y=0$, we apply a frictional boundary condition. In principle, this means that we adopt the linearization in an approximate sense. i.e. small deviations from linearity can cause a difference in shear stress between $y=0$ and $y=H$. The mathematical formulation is also equivalent to treating the y-coordinate as quasi-static, where the relevant difference in shear stress reduces to the difference between the shear stress at $y \rightarrow 0$ (which equals the stress at $y=H$), and the kinetic shear stress at the boundary ($y=0$). We choose an Amontons-Coulomb friction law $f_f = \hat \sigma_{x}(x,0,t)$ at the interface, which can be written as
\begin{align}
    f_f 
    \begin{cases}
        \leq \mu_s \sigma_n & \text{if } \dot u = 0 \\
        = \mu_k \sigma_n & \text{if } |\dot u| > 0
    \end{cases}
    \label{eq:ff_app}
\end{align}
where $\sigma_n$ is the normal stress (which is equal to $\sigma_{yy}$), with the additional criteria that a transition from static to dynamic friction occurs when the local shear stress is greater than $\mu_s \sigma_n$, and a transition from dynamic to static friction occurs when the local velocity reaches zero. In the following, we will write $u_x(x,0,t)$ as $u$. 

\section{Dimensionless formulation \label{app:eom_nondim}}
For the rest of the document, and in the main text, we drop the subscript $_x$. We start with the equation of motion using a displacement boundary condition at $y=H$. Following the approach from \citet{amundsen2015steady}, we assume an initial displacement field $u(x,0,0)$, and a constant displacement $u(x,H,0)$. We then change the coordinate system $u'= u(x,0,t)-u(x,0,0)$, which allows us to rewrite the initial displacement to an initial shear stress $\tau$:
\begin{equation}
    \ddot{u}' = \frac{\lambda + 2G}{\rho} \frac{\partial^2 u'}{\partial x^2} - \frac{2G}{\rho H^2}u' - \frac{2}{\rho H} \hat \sigma_{x}(x,0,t) + \frac{2}{\rho H} \tau
\end{equation}
We then define the dimensionless variables $\bar{u} = \frac{u'}{U}$, $\bar{t}= \frac{t}{T}$ and $\bar x = \frac{x}{X}$ so that
\begin{equation}
\ddot{\bar u} U/T^2 = \frac{\lambda + 2G}{\rho} \frac{U}{X^2} \frac{\partial^2 \bar u}{\partial \bar x^2} - \frac{2 G U}{\rho H^2} \bar u - \frac{2}{\rho H}\hat \sigma_{x}(x,0,t) + \frac{2}{\rho H} \tau,
\end{equation}
where the derivative is now taken with respect to $\bar t$ and $\bar x$. We can further manipulate the expression to arrive at
\begin{equation}
\ddot{\bar u} = \frac{\lambda + 2G}{\rho} \frac{T^2}{X^2} \frac{\partial^2 \bar u}{\partial \bar x^2} - \frac{2 G T^2}{\rho H^2} \bar u - \frac{T^2}{U} \frac{2}{\rho H}\hat \sigma_{x}(x,0,t) + \frac{T^2}{U} \frac{2}{\rho H} \tau,
\end{equation}
Next, we reduce the number of parameters by selecting $\frac{\lambda + 2G}{\rho} \frac{T^2}{X^2} = 1$, which leads to $T = X \sqrt{\frac{\rho}{\lambda + 2G}}$. This automatically obeys dimensionless wave speed equal to 1 in the limit of small $\frac{2 G T^2}{\rho H^2}$;
\begin{equation}
\bar v_s = \sqrt{\frac{\lambda + 2G}{\rho}} \frac{T}{X} = \sqrt{\frac{\lambda + 2G}{\rho}} \sqrt{\frac{\rho}{\lambda + 2G}} = 1.
\end{equation}
Choosing $U = 2 H \frac{\mu_\text{s} p - \mu_\text{k} \sigma_n}{\lambda + 2G}$ leads to
\begin{align}
\frac{T^2}{\rho U H}(\tau - f_{f}) = \frac{X^2}{H^2}\frac{ \tau - \hat \sigma_{x}}{\mu_\text{s} \sigma_n - \mu_\text{k} \sigma_n}.
\end{align}
If we now select the length scale $X=H$, we obtain the familiar parameter \cite{amundsen2015steady,thogersen2019minimal}
\begin{equation}
\frac{T^2}{\rho U}(\tau - \hat \sigma_{x}) = \frac{ \tau/\sigma_n \mp \mu_\text{k}}{\mu_\text{s} - \mu_\text{k}} \equiv \bar \tau^\pm
\end{equation}
which applies when the interface is sliding, and where $\pm$ corresponds to the sign of the velocity. In this paper only positive velocities appear, which means we can simplify to $\bar \tau \equiv \bar \tau^+$. We can then define the non-dimensional equation of motion
\begin{equation}
\ddot{\bar u} = \frac{\partial^2 \bar u}{\partial \bar x^2} - \bar \gamma {\bar u} + \bar \tau.
\end{equation}
where 
\begin{equation}
\bar \gamma = \frac{2 G T^2}{\rho H^2} = \frac{2 G H^2 \rho}{\rho H^2 (\lambda + 2G)} = \frac{2 G}{ (\lambda + 2G)}.
\end{equation}
To remove oscillations at the grid scale we introduce a damping term $\bar \beta$
\begin{equation}
\ddot{\bar u} = \frac{\partial^2 \bar u}{\partial \bar x^2} - \bar \gamma {\bar u} + \bar \tau - \bar \beta \frac{\partial ^2 \dot {\bar u}}{\partial \bar x^2}.
\label{eq:eom_app}
\end{equation}

Equation \ref{eq:eom_app} applies for the sliding part of the interface, and dynamic friction is included through $\bar \tau$. The static friction threshold in dimensionless units reduces to
\begin{equation}
    |\frac{\partial^2 \bar u}{\partial \bar x^2} - \bar \gamma \bar u + \bar \tau| \geq 1
\end{equation}

The same approach can be carried out using the equation with stress boundary condition, and we define the non-dimensional equation of motion
\begin{equation}
\ddot{\bar u} = \frac{\partial^2 \bar u}{\partial \bar x^2} + \bar \tau - \bar \beta \frac{\partial ^2 \dot {\bar u}}{\partial \bar x^2}.
\end{equation}
where the dimensionless scaling parameters are the same as before apart from $U$ which changes by a factor two due to the different way of integrating the thin layer for the two different boundary conditions: $U = H \frac{\mu_\text{s} p - \mu_\text{k} p}{\lambda + 2G}$. 

In the manuscript, we combine these two boundary conditions in a single equation of motion
\begin{equation}
\ddot{\bar u} = \frac{\partial^2 \bar u}{\partial \bar x^2} - \Gamma \bar \gamma {\bar u} + \bar \tau - \bar \beta \frac{\partial ^2 \dot {\bar u}}{\partial \bar x^2}.
\label{eq:eom_final}
\end{equation}
where $\Gamma = 1$ corresponds to displacement boundary condition, and $\Gamma = 0$ corresponds to stress boundary condition. It is worth noting that the finite difference scheme of the dimensionless equation of motion derived here is identical to the classical Burridge-Knopoff spring block model with leaf springs (see e.g. \citet{heaton1990evidence}) under the assumption that the discretization $\Delta x$ equals the thickness of the system $\Delta x = H$.

\section{Analytical solution for steady state slip pulse with displacement boundary condition (\texorpdfstring{$\bar \beta = 0$}{}) \label{app:analytical}}
To obtain the steady state solution for a slip pulse, we first assume a homogeneous stress profile $\tau(x)=\tau_0$ and define
\begin{equation}
\bar\tau_0 \equiv \frac{ \tau_0/\sigma_n \mp \mu_\text{k}}{\mu_\text{s} - \mu_\text{k}}.
\end{equation}
Next, we solve
\begin{equation}
\ddot {\bar u} = \frac{\partial^2 \bar u}{\partial \bar \chi ^2} + \bar \tau_0 - \bar \gamma \bar u
\end{equation}
where $\bar \chi$ is a co-moving coordinate with units of $\bar x$ where $\bar \chi = 0$ corresponds to the position of the maximum slip speed inside the pulse. We assume $\dot{\bar u}(-\bar W/2) = \dot{\bar u}(\bar W/2) = 0$, $\bar u(-\bar W/2) = \Delta \bar \tau / \bar \gamma$, and $\bar u (\bar W/2) = 0$, where $\bar W = \frac{W}{H}$ is the dimensionless pulse width. For a constant front speed $\bar v_c$ we can relate $\bar {u}$ to $\dot {\bar u}$ and $\ddot {\bar u}$:
\begin{equation}
\dot{\bar u} = -\bar v_c \frac{\partial \bar u}{\partial \bar \chi}
\end{equation}
where the minus sign arises from the assumption of propagation in the positive direction. Then,
\begin{equation}
\ddot{\bar u} = \bar v_c^2 \frac{\partial^2\bar u}{\partial \bar \chi^2},
\end{equation}
which leads to
\begin{equation}
\bar v_c^2 \frac{\partial^2\bar u}{\partial \bar \chi^2} = \frac{\partial^2 \bar u}{\partial \bar \chi ^2} + \bar \tau_0 - \bar \gamma \bar u,
\end{equation}
which can be rewritten as
\begin{equation}
\frac{\partial^2\bar u}{\partial \bar \chi^2}\left ( \bar v_c^2 - 1 \right ) + \bar \gamma \bar u - \bar \tau_0 = 0.
\end{equation}
The general solution is 
\begin{equation}
\bar u (\bar \chi) = c_1 \sin \left ( \frac{\sqrt{\bar \gamma} \bar x}{\sqrt{\bar v_c^2 - 1}} \right ) + c_2 \cos \left ( \frac{\sqrt{\bar \gamma} \bar \chi}{\sqrt{\bar v_c^2 - 1}} \right ) + \frac{\bar \tau_0}{\bar \gamma}
\end{equation}

\begin{figure*}
    \centering
    \includegraphics{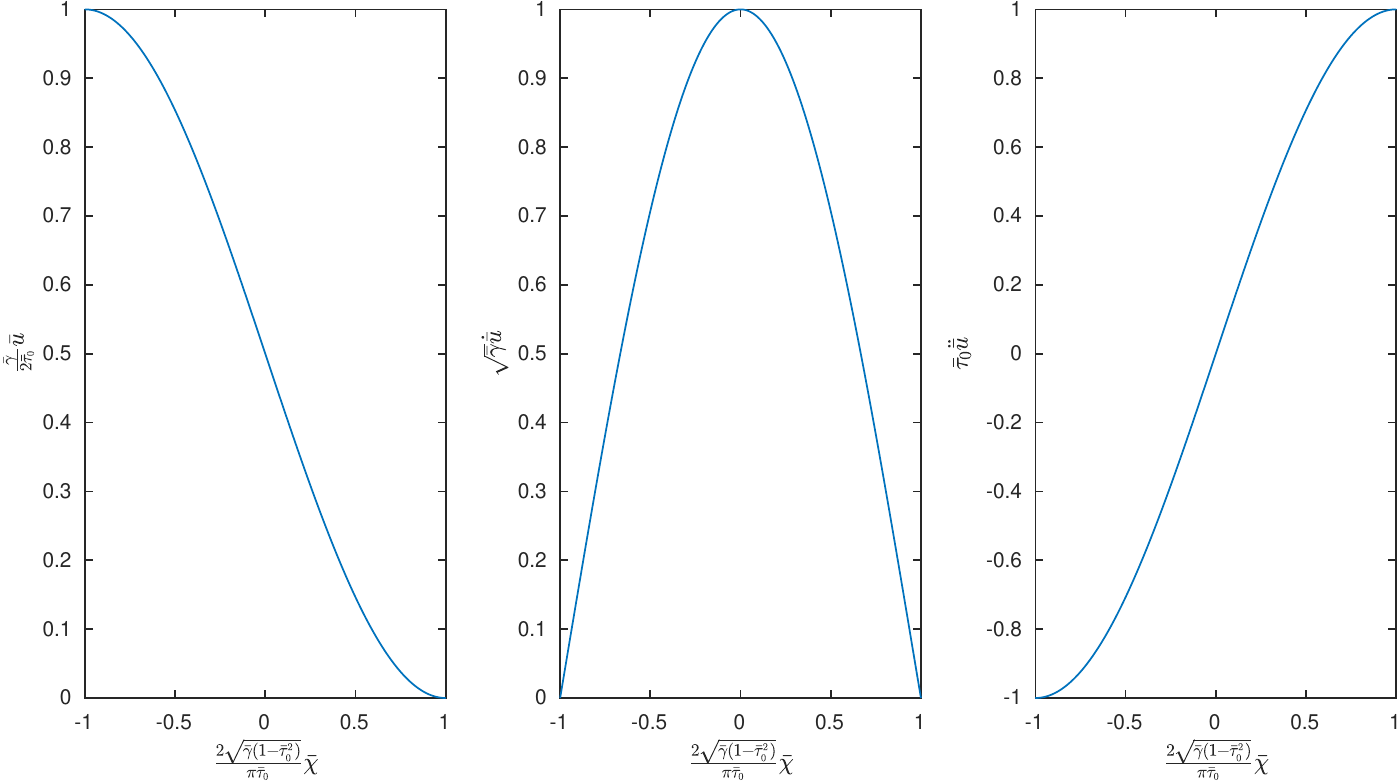}
    \caption{Steady state solution of slip, slip velocity, and acceleration for $\Gamma = 1$.
    \emph{Left}: Non-dimensional slip in the pulse in steady state.
    \emph{Middle}: Non-dimensional slip velocity in the pulse in steady state.
    \emph{Right}: Non-dimensional acceleration inside the propagating pulse in steady state.}
    \label{fig:steady_state_app}
\end{figure*}

\begin{figure*}
    \centering
    \includegraphics{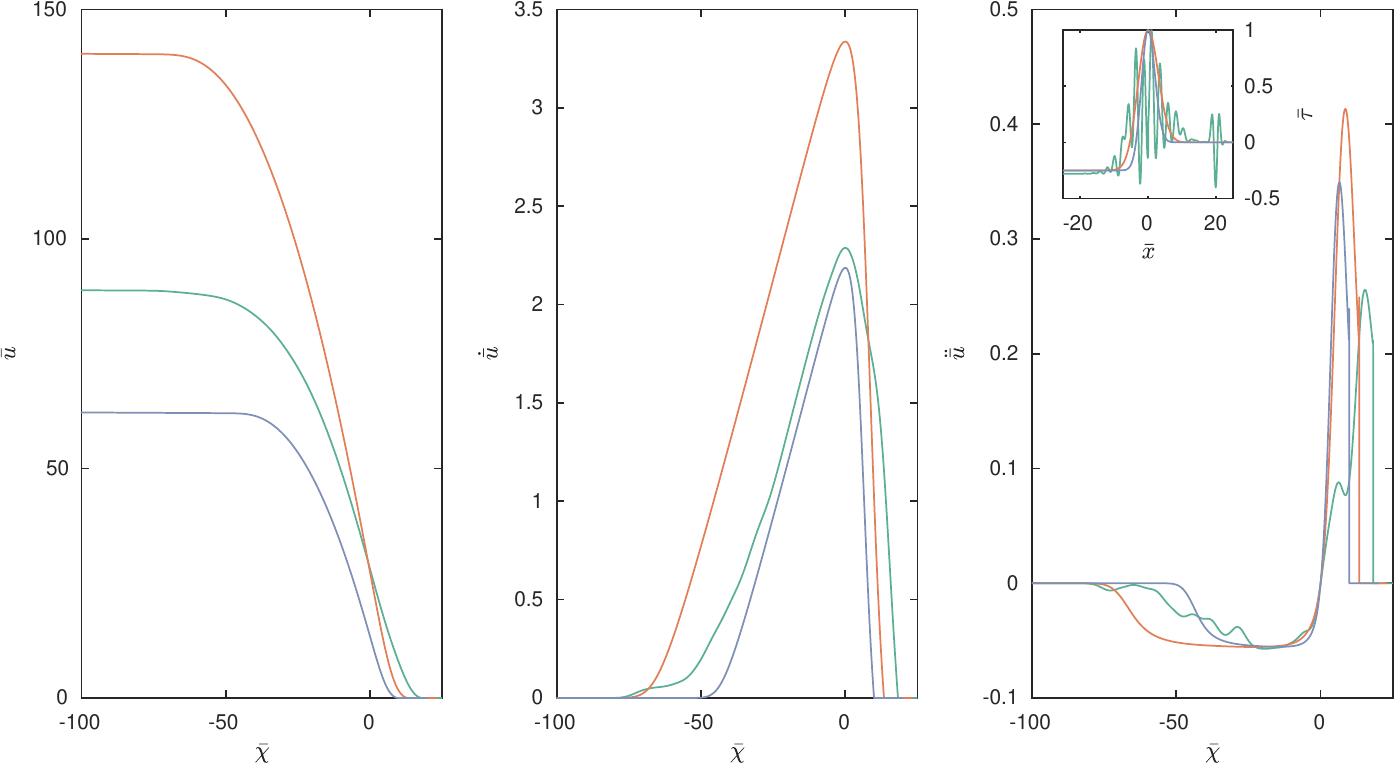}
    \caption{Steady state slip, slip velocity and acceleration/stress for a selection of pulses at $\Gamma = 0$ with prestress shown in the inset.
    \emph{Left}: Non-dimensional slip in the pulse in steady state.
    \emph{Middle}: Non-dimensional slip velocity in the pulse in steady state.
    \emph{Right}: Non-dimensional acceleration inside the propagating pulse in steady state.}
    \label{fig:steady_state_zerogamma_app}
\end{figure*}

To find the constants $c_1$, $c_2$, as well as the pulse width $\bar W$ and the stress drop $\Delta \bar \tau$, we use the following boundary conditions: $\bar u (\bar W/2) = 0$, $\bar u(-\bar W/2) = \frac{\Delta \bar \tau_0}{\bar \gamma}$, where $\Delta \bar \tau$ is the stress drop. $\dot {\bar u}(\bar W/2) = 0$, $\dot {\bar u}(-\bar W/2) = 0$. In addition, we set $\bar v_c = 1/\sqrt{1-\bar \tau^2}$ which is exact in the limit $\bar \gamma \rightarrow 0$ \cite{amundsen2015steady}, but is also a good approximation for $\bar \gamma \lesssim 1$ as seen in \figurename~\ref{fig:steady_state} in the manuscript.
This gives us the solution
\begin{equation}
\bar u(\bar \chi) = \frac{\bar \tau_0}{\bar \gamma} \left ( 1 - \sin(\frac{\sqrt{\bar \gamma (1 - \bar \tau_0^2)}}{\bar \tau_0}\bar \chi) \right )
\end{equation}
for $\bar \chi \in [-\bar W/2, \bar W/2]$, where $\bar W$ is the pulse width
\begin{equation}
\bar W = \pi \sqrt{\frac{\bar v_c^2 - 1}{\bar \gamma}} = \frac{\pi \bar \tau_0}{\sqrt{\bar \gamma (1-\bar \tau_0^2)} }
\end{equation}
The slip speed inside the pulse is
\begin{align}
\dot{\bar u}(\bar \chi) = \frac{1}{\sqrt{\bar \gamma}}
\cos \left ( \frac{\sqrt{\bar \gamma (1-\bar \tau_0^2)}}{\bar \tau_0} \bar \chi \right ) 
\end{align}
where the maximum slip speed is
\begin{equation}
\dot{\bar u}_\text{max} = \frac{1}{\sqrt{\gamma}}.
\end{equation}
The slip acceleration is
\begin{equation}
    \ddot{\bar u}(\bar \chi) = \frac{1}{\bar \tau_0} \sin \left ( \frac{ \sqrt{\bar \gamma (1 - \bar \tau_0^2)} }{\bar \tau_0} \bar \chi \right )
\end{equation}
In addition, the stress drop is given by
\begin{equation}
\Delta \tau = 2\bar \tau_0
\end{equation}
which can also be understood from symmetry arguments. \figurename~\ref{fig:steady_state_app} summarizes the pulse solution in steady state. \figurename~\ref{fig:steady_state_zerogamma_app} shows pulse solutions for $\Gamma = 0$ expanding on the data shown in the main text.

\section{Scaling relationships for systematic variation of the damping parameter \texorpdfstring{$\bar \beta$}{} \label{app:scaling_damping}}
\begin{figure*}[b]
    \centering
    \includegraphics{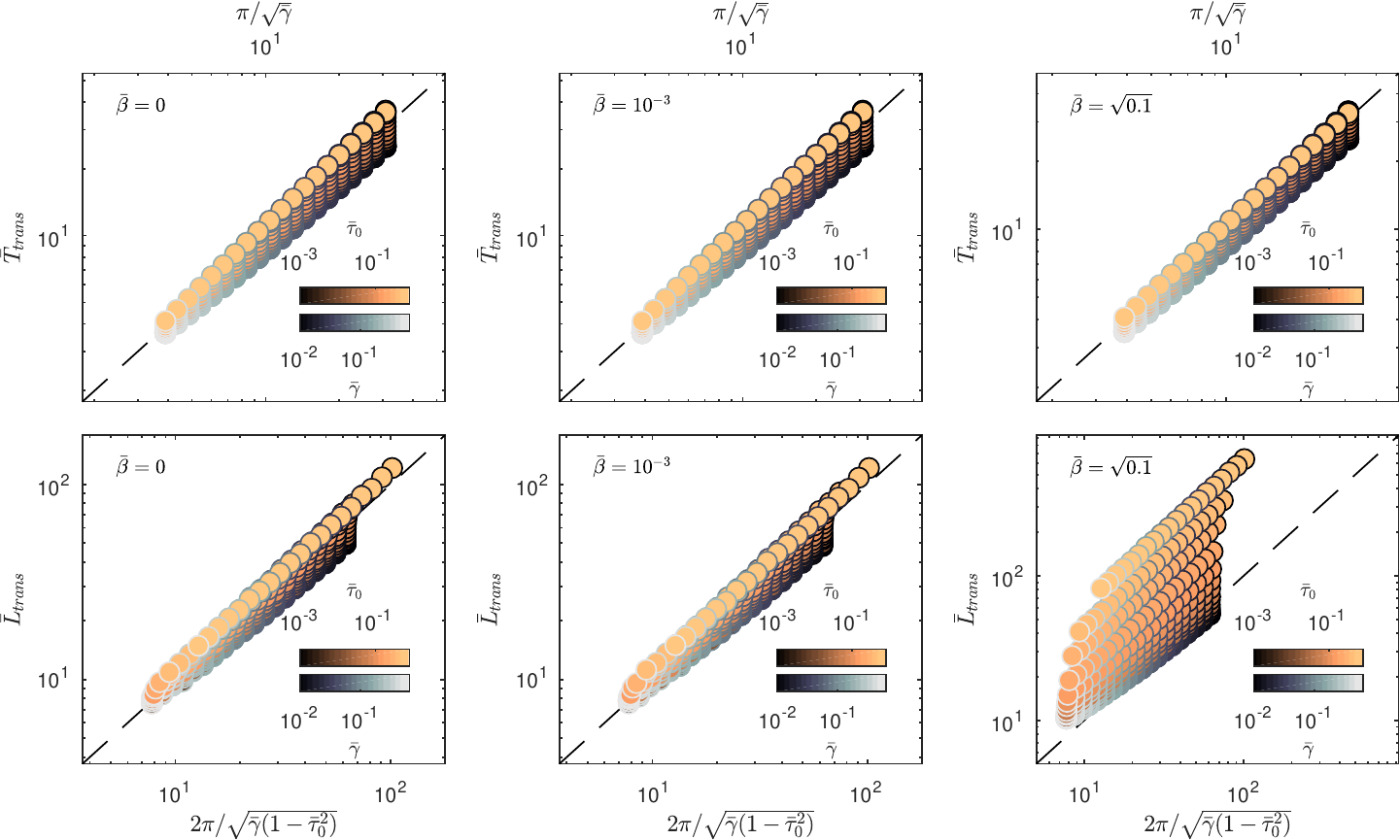}
    \caption{Scaling relationships for the transition from crack-like to pulse-like rupture for a variety of damping parameters $\bar \beta$ for $\Gamma = 1$.}
    \label{fig:collapse_nonZeroGamma}
\end{figure*}

\begin{figure*}[b]
    \centering
    \includegraphics{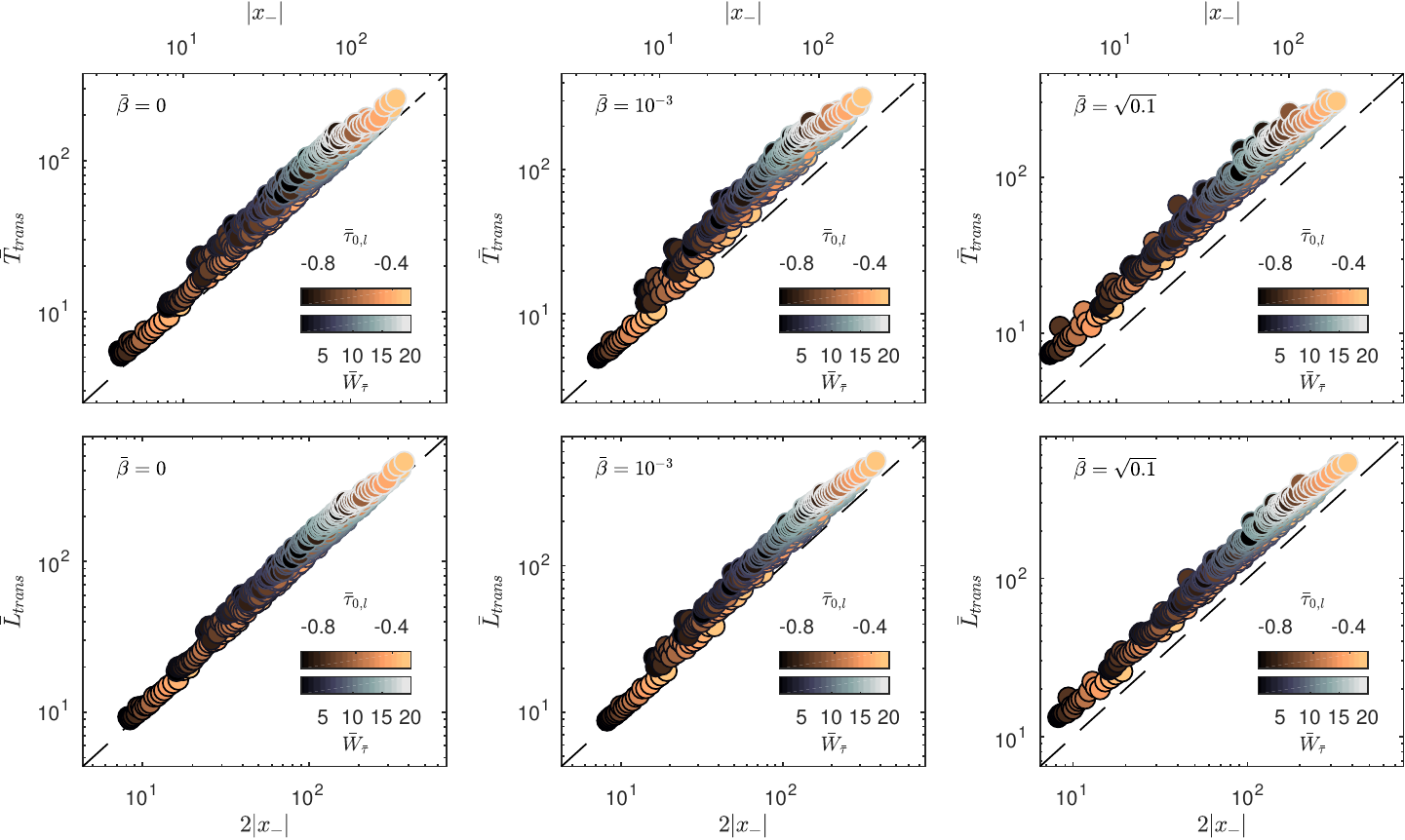}
    \caption{Scaling relationships for the transition from crack-like to pulse-like rupture for a variety of damping parameters $\bar \beta$ for $\Gamma = 0$}
    \label{fig:collapse_ZeroGamma}
\end{figure*}

In the main text, we present a set of scaling relationships for the transition from crack-like to pulse like rupture. \figurename~\ref{fig:collapse_nonZeroGamma} shows how the data collapse for $\Gamma = 1$ is affected by the damping parameter $\bar \beta$. \figurename~\ref{fig:collapse_ZeroGamma} shows how the data collapse for $\Gamma = 0$ is affected by the damping parameter $\bar \beta$. 


\end{document}